# Transboundary Secondary Organic Aerosol in Western Japan: An Observed Limitation of the $f_{44}$ Oxidation Indicator

Satoshi Irei,[1,¶,*] Akinori Takami,[1] Yasuhiro Sadanaga,[2] Takao Miyoshi,[1] Tekemitsu Arakaki,[3] Kei Sato,[1] Naoki Kaneyasu,[4] Hiroshi Bandow,[2] and Shiro Hatakeyama[5]

[1]National Institute for Environmental Studies, 16-2 Onogawa, Tsukuba, Ibaraki 305-8506, Japan

[2]Department of Applied Chemistry, Graduate School of Engineering, Osaka Prefecture University, 1-1 Gakuencho, Naka-ku, Sakai, Osaka 599-8531, Japan

[3]Department of Chemistry, Biology, and Marine Science, University of the Ryukyus, 1 Senbaru, Nishihara, Okinawa 903-0213, Japan

[4]Agricultural Department, Tokyo University of Agriculture and Technology, 3-5-8 Saiwai-cho, Fuchu, Tokyo 183-8509, Japan

[*]Corresponding author: Satoshi Irei, Department of Chemistry, Biology, and Marine Science, University of the Ryukyus, 1 Senbaru, Nishihara, Okinawa 903-0213, Japan (fax: +81-98-895-8386; e-mail: satoshi.irei at gmail.com)

[¶]Present address: Department of Chemistry, Biology and Marine Science, Faculty of Science, University of the Ryukyus, 1 Senbaru, Nishihara, Okinawa 903-0213, Japan

**Abstract:** To obtain evidence for secondary organic aerosol formation during the long-range transport of air masses over the East China Sea, we conducted field measurements in March 2012 at the Fukue atmospheric monitoring station, Nagasaki, in western Japan. The relative abundance of $m/z$ 44 in fine organic aerosol ($f_{44}$) was measured by an Aerodyne aerosol chemical speciation monitor. The stable carbon isotope ratio ($\delta^{13}C$) of low-volatile water-soluble organic carbon (LV-WSOC) in the daily filter samples of total suspended particulate matter was also analyzed using an elemental-analyzer coupled with an isotope ratio mass spectrometer. Additionally, in situ measurements of $NO_x$ and $NO_y$ were performed using $NO_x$ and $NO_y$ analyzers. The measurements showed that, unlike the systematic trends observed in a previous field study, a scatter plot for $\delta^{13}C$ of LV-WSOC versus $f_{44}$ indicated a random variation. Comparison of $f_{44}$ with the photochemical age estimated by the $NO_x/NO_y$ ratio revealed that the $f_{44}$ values distributing around 0.2 likely reached a saturation level already and the $f_{44}$ values were significantly lower than the observed $f_{44}$ (~0.3) at Hedo in the previous study. These findings imply that the saturation point of $f_{44}$, and the use of $f_{44}$ as an oxidation indicator, is case dependent.

**Introduction**

A quantitative understanding of atmospheric secondary organic aerosol (SOA), one of the least understood areas of atmospheric chemistry, is urgently required for a more accurate estimation of radiative forcing[1] and to better understand its association with adverse health effects.[2] To understand the formation of SOA, an evaluation of the extent of oxidation progress is necessary. Recently, we reported the results of field SOA studies at rural sites in western Japan, using stable carbon isotope and aerosol mass spectrometric techniques as a metric for oxidation processing.[3] The study reported novel findings, with two different systematic trends identified from a plot of the stable carbon isotope ratio ($\delta^{13}$C) of low-volatile water-soluble organic carbon (LV-WSOC) in filter samples, as a function of the relative abundance of $m/z$ 44 ($CO_2^+$, fragment ion of carboxylic acids) in organic aerosol (OA) mass spectra measured by aerosol mass spectrometers (often denoted as $f_{44}$). Furthermore, a proportional increase of $f_{44}$ with photochemical age ($t$[OH]) estimated by $NO_x/NO_y$ ratios was found in the study, validating the usefulness of $f_{44}$ and $\delta^{13}$C as oxidation indicators. Controversially, it has been reported that $f_{44}$ sometimes has a good association with $t$[OH],[4] but on other occasions does not.[5-7] Further evaluation is needed because the study period was only 11 days. In this paper, we report an observed limitation of the $f_{44}$ indicator in a different case study at Fukue Island, a rural site in western Japan.

**Materials and Method**

A field study was conducted from 8 to 18 March 2012 at the Fukue atmospheric monitoring station (32.8°N, 128.7°E), shown in Figure 1. During the studies, total suspended particulate matter was collected every 24 h (noon to noon) on a pre-baked (773 K for 12 h) quartz fiber filter (8 × 10-inch Tissuquartz™ filters: Pall Corp., Port Washington, NY, USA) using a high-volume air sampler (HV-1000: Sibata Corp., Saitama, Japan). A quarter of each filter sample was used for the LV-WSOC analysis. The details of the analytical method are described elsewhere.[1] Briefly, LV-WSOC on a filter sample was extracted three times by ultrasonic agitation with ultra-pure water (Wako Pure Chemical Industries, Ltd., Osaka, Japan). The extracts were filtered using a 0.45-µm polytetrafluoroethylene (PTFE) syringe filter (PURADISC 25TF: Whatman Japan K.K., Tokyo, Japan) and were then combined. The volume of the combined extract was reduced to ~0.1 mL using a rotary evaporator (R-205 and B-490: Nihon Büchi K.K., Tokyo, Japan), followed by further reduction in a gentle flow of 99.99995% pure nitrogen (Tomoe Shokai, Tokyo, Japan) in a pre-weighed conical vial (Mini-vial: GL Sciences Inc., Tokyo, Japan). The final volume of the concentrated extract was determined by weighing the extract in the vial. For quantitative elemental and isotope ratio analysis, an aliquot of the concentrated extract was pipetted into a 0.15-mL tin cup (Ludi Swiss AG, Flawil, Switzerland). The extract in the cup was then dried under a gentle flow of pure nitrogen. A drop of 0.01 M hydrochloric acid (Wako Pure Chemical Industries) was spiked into the dried sample for the removal of carbonate, and the sample was re-dried. The samples prepared in this manner were analyzed using an elemental analyzer (Flash 2000: Thermo Scientific, Waltham, MA, USA) coupled with an open-split interfaced (Conflo IV: Thermo Scientific) isotope ratio mass spectrometer (Delta V Advantage: Thermo Scientific) for determining the carbon mass and $\delta^{13}C$ value. Here, $\delta^{13}C$ is defined as follows:

$$\delta^{13}C = \left[ \frac{(\frac{^{13}C}{^{12}C})_{sample}}{(\frac{^{13}C}{^{12}C})_{reference}} - 1 \right],$$

where $(^{13}C/^{12}C)_{sample}$ and $(^{13}C/^{12}C)_{reference}$ are the $^{13}C/^{12}C$ atomic ratios for the sample and the reference (the Vienna Pee Dee Belemnite), respectively.

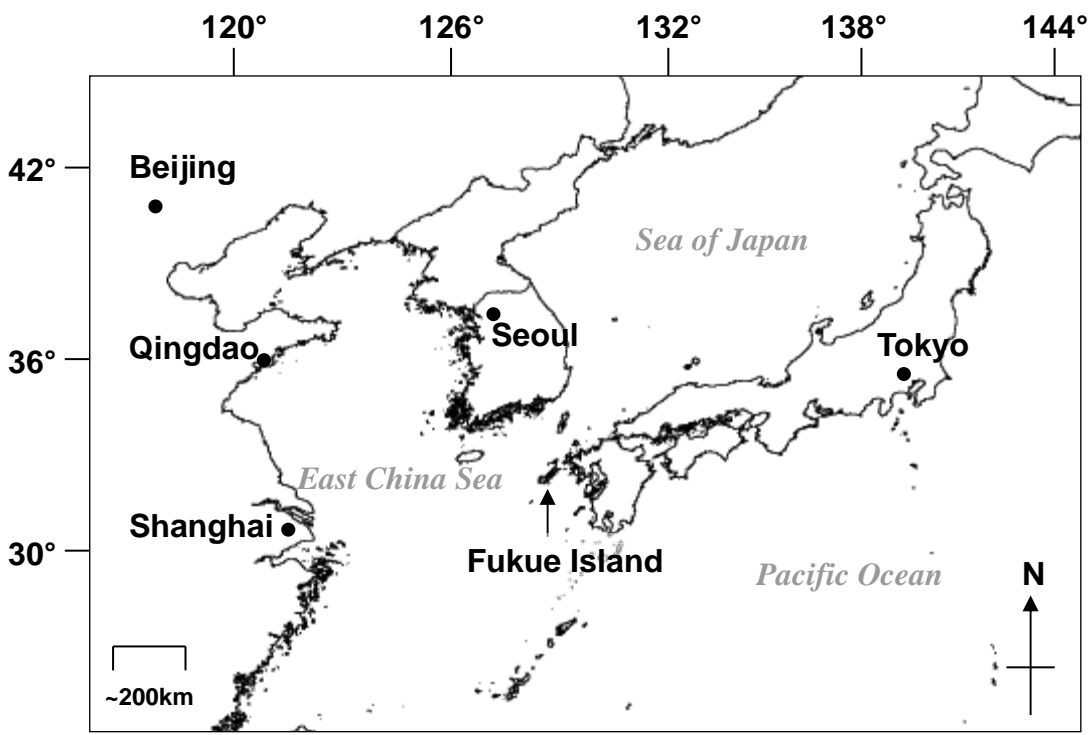

**Figure 1. Map showing the location of measurement site.**

During the study period, an aerosol chemical speciation monitor (ACSM: Aerodyne Research Inc., Billerica, MA, USA) was also used to measure the chemical composition of fine aerosol < 1 µm aerodynamic diameter (i.e., $PM_{1.0}$). The instrument is similar to an aerosol mass spectrometer,[8] which uses a unique method for the calculation of chemical species concentration.[9] The detail of the ACSM are described elsewhere.[10] Briefly, the flash vaporizer of the ACSM was set to 873 K and the instrument scanned the ions of evaporated substances from *m/z* 1 to 150 to determine the atmospheric concentrations of organics, sulfate, nitrate, and ammonium. The instrument was calibrated with 350-nm dried ammonium nitrate particles at the beginning of the quantitative analysis. The response factor using this reagent was $2.3 \times 10^{-11}$ A m$^3$ µg$^{-1}$. To verify the concentrations determined by the ACSM, 24 h averaged concentrations of sulfate measured by the ACSM were compared with daily average concentrations of non-sea-salt sulfate (= total sulfate concentration – $0.251 \times [Na^+]$) obtained from the total suspended particulate matter. This comparison was used to establish a collection efficiency of 0.72 for the ACSM measurements. Note that for conciseness, only the concentrations of OA and *m/z* 44 are presented here and discussed in the following sections.

The mixing ratios of NO and $NO_2$ (i.e., $NO_x$) and of total odd nitrogen (i.e., $NO_y$) were also measured during the study using $NO_x$ and $NO_y$ analyzers that were developed from commercially available $NO_x$ analyzers (Model 42 i-TL: Thermo Scientific) by Sadanaga et al.[11] and Yuba et al.[12] The $NO_x$ and $NO_y$ measurements were valuable because the $NO_x/NO_y$ ratio provides another atmospheric oxidation indicator, *t*[OH]. The *t*[OH] is derived by the second order rate law, assuming that an irreversible conversion of $NO_2$ to $HNO_3$ is the major sink of $NO_x$:[6, 13]

$$t[\text{OH}] = -\frac{1}{k_{NO_2}} \ln \frac{[NO_x]}{[NO_y]} \qquad (1)$$

where *t*, [OH], [$NO_x$], [$NO_y$], and $k_{NO2}$ are the reaction time, the average concentration of OH radical during the reaction, the concentrations of $NO_x$ and $NO_y$ at time *t*, and the second order rate constant for the conversion of $NO_2$ to $HNO_3$ by oxidation with OH radical, respectively. We used $8.7 \times 10^{-12}$ cm$^3$ molecule$^{-1}$ s$^{-1}$ as the $k_{NO2}$ value at 300 K and 1 atm[14] for the calculation of *t*[OH]. No measurement results for $NO_y$ were available between 7 and 10 March due to maintenance of the instrument.

**Results and Discussion**
**Organics Measured by ACSM**

During the study the *m/z* 44 concentrations varied in a similar way to the OA concentrations (Figure 2), indicating that the OA was strongly associated with carboxylic acid. The average concentrations and standard deviations (SD) of OA and *m/z* 44 were 4.7 ± 3.0 µg m$^{-3}$ and 0.85 ± 0.52 µg m$^{-3}$, respectively. Compared to the average concentrations observed at the same location in the 2010 study (2.5 ± 1.6 µg m$^{-3}$ and 0.33 ± 0.23 µg m$^{-3}$ for the OA and the *m/z* 44, respectively), the average values in this study were almost two orders of magnitude higher. Back trajectories of air masses indicated that the air masses, with the exception of those arriving on 17 and 18 March, passed close to north-eastern China and/or the Korean peninsula (Figure 3), implying that they probably had a transboundary origin.

We ran a positive matrix factorization (PMF) analysis on the organic mass spectra to better understand its composition. In contrast to the field studies in 2010, we found that one factorial analysis provided the most feasible solution with respect to the mass spectra patterns (Figure 4). Compared to the typically observed mass spectra of factorial components (AMS Spectral Database by Ulbrich et al., http://cires.colorado.edu/jimenez-group/AMSsd/), the extracted component coincided with the spectra for low-volatile oxygenated organic aerosol (LV-OOA). An elemental analysis of LV-OOA conducted by Zhang et al.[15] confirmed the ratio of organic aerosol mass to organic carbon mass (OM/OC) to be 4.3 *µg* per *µg* carbon (*µg*C$^{-1}$), which is similar to the ratios of fulvic and humic acids (4.2-4.6), known as humic like substances (HULIS), according to the reference mass spectra for these compounds as cited in the database. The OM/OC ratio was slightly higher than that observed in the 2010 study (3.6 *µg* *µg*C$^{-1}$). From the current data set, we cannot conclude that the difference was negligible or was caused by the OA composition.

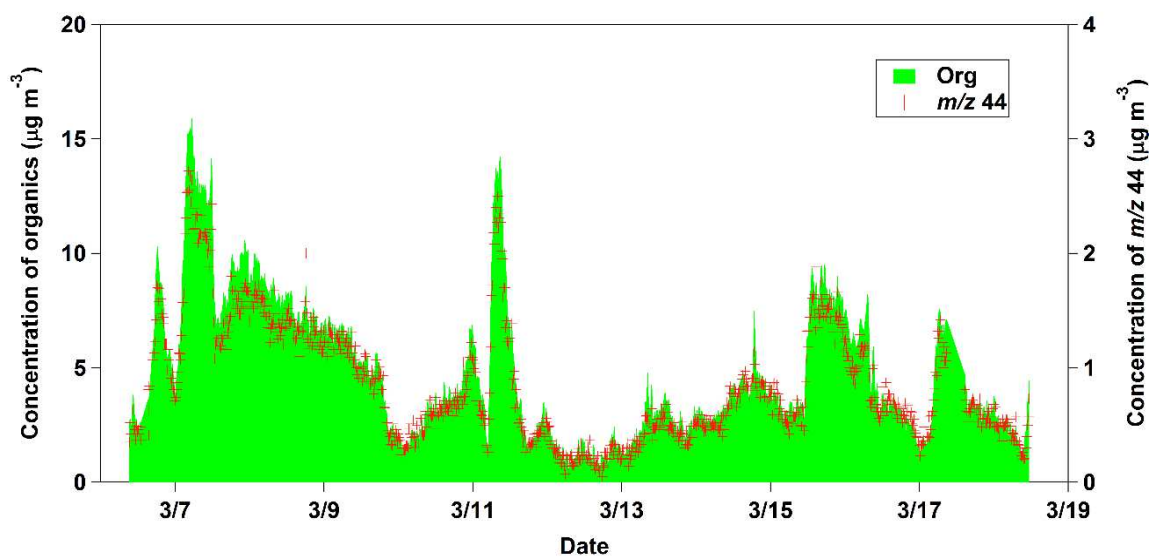

**Figure 2. Time series of the organic aerosol concentration and its *m/z* 44 component measured by ACSM at Fukue.**

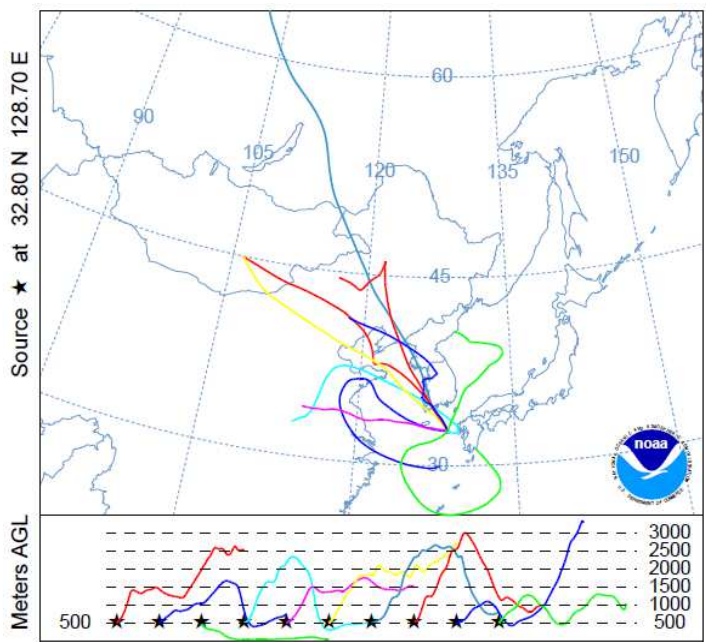

**Figure 3. 72 hour back trajectories of 500 m AGL air masses arriving at Fukue during the study period. The trajectories were drawn every 24 hour, and the star marks from right to left correspond to the trajectories from March 8th to 18th, 2012.**

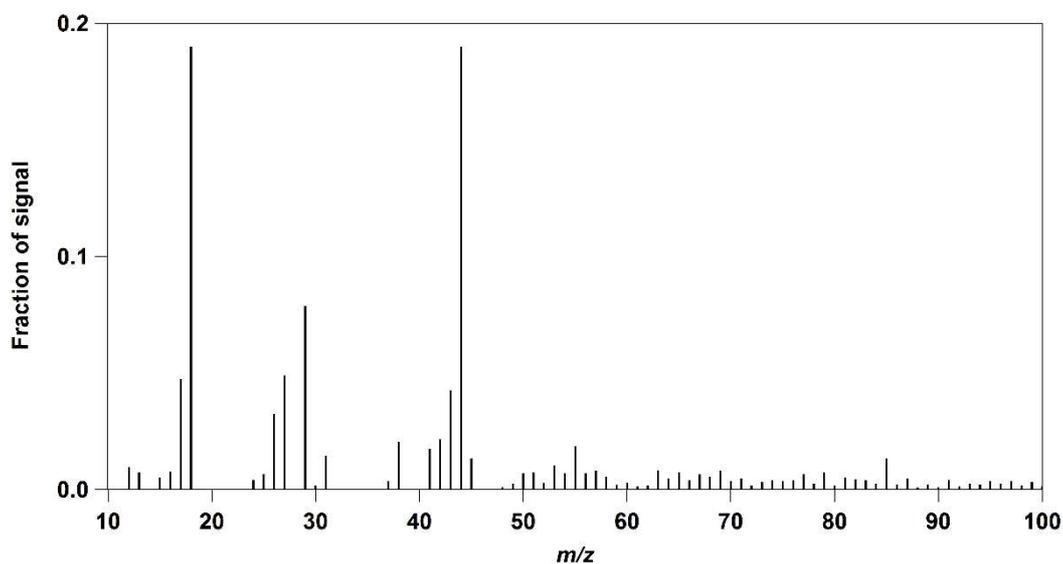

**Figure 4. Mass spectra of LV-OOA extracted by one factorial PMF analysis.**

**Filter sample analysis for LV-WSOC**

During the study period, the daily averaged LV-WSOC concentrations ranged between 0.9 and 5.1 $\mu g$C m$^{-3}$, with a mean ± SD of 2.0 ± 1.0 $\mu g$C m$^{-3}$ (Figure 5). These values were nearly half of those observed in the 2010 field study at Fukue (from 0.4 to 2.1 $\mu g$C m$^{-3}$ with a mean of 1.1 ± 0.5 $\mu g$C m$^{-3}$). A comparison of the OM/OC ratio and the ratio of the OA measured by ACSM to the LV-WSOC (OA/LV-WSOC), gives an indication of the difference between the fine aerosol measured by ACSM and the particles collected on filters. Based on the average values, the OA/LV-WSOC was 2.3 $\mu g\ \mu gC^{-1}$. This ratio is nearly half of the OM/OC ratio of 4.3 $\mu g\ \mu gC^{-1}$. Because the PMF analysis already revealed the composition of OA with a single factor, this difference implies that half of the LV-WSOC was distributed in particle sizes larger than $PM_{1.0}$, which is consistent with the results of the previous study.

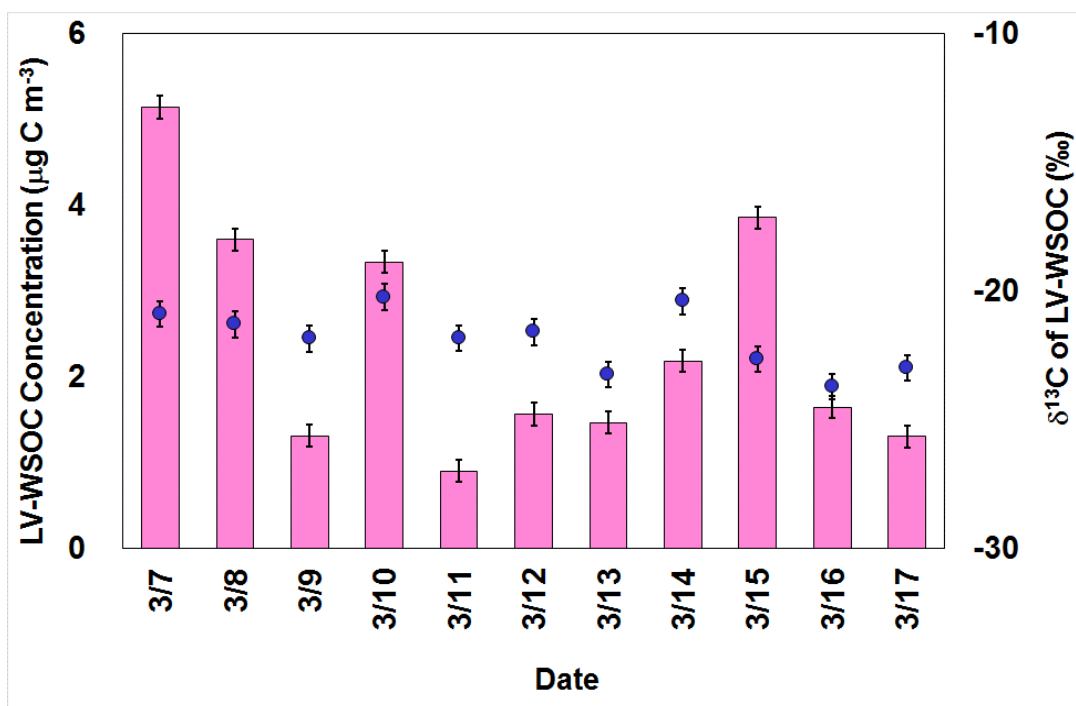

**Figure 5. Time series plot of the concentration and $\delta^{13}C$ of LV-WSOC observed at Fukue stations.**

The $\delta^{13}C$ values of LV-WSOC at Fukue were between -19.7 and -22.8 ‰, with a mean ± SD of -21.1 ± 1.1 ‰. In the previous study, the $\delta^{13}C$ of LV-WSOC varied from -19.3 to -29.4 ‰, with a mean of -22.0 ± 2.0 ‰. Although the variation ranges observed during the two different field studies were similar, the mean $\delta^{13}C$ value in this study was approximately 1 ‰ heavier at Fukue.

**Mixing ratio of $NO_x$ and $NO_y$, and estimated $t$[OH]**

A time series plot demonstrated that $NO_x$ ranged from 0.60 to 8.9 ppbv, with a mean ± standard deviation (SD) of 1.4 ± 0.8 ppbv (Figure 6). The $NO_y$ ranged from 1.4 to 20.4 ppbv, with a mean of 5 ± 3 ppbv. Their average mixing ratios in the previous study (1.5 and 3ppbv, respectively) were of similar magnitude, and in most cases the majority of $NO_x$ was $NO_2$. Although the mixing ratios of $NO_y$ between 7 and 10 March are not known, its variation from 11 to 17 March roughly resembled the variation of LV-WSOC and OA, indicating that the origin of OA, LV-WSOC, and/or its precursor, and $NO_y$ are likely the same. Based on these mixing ratios, the calculated $t$[OH] ranged from $4.2\times10^6$ to $7.7\times10^7$ h molecule $cm^{-3}$, with a mean of $4\times10^7 \pm 1\times10^7$ h molecule $cm^{-3}$.

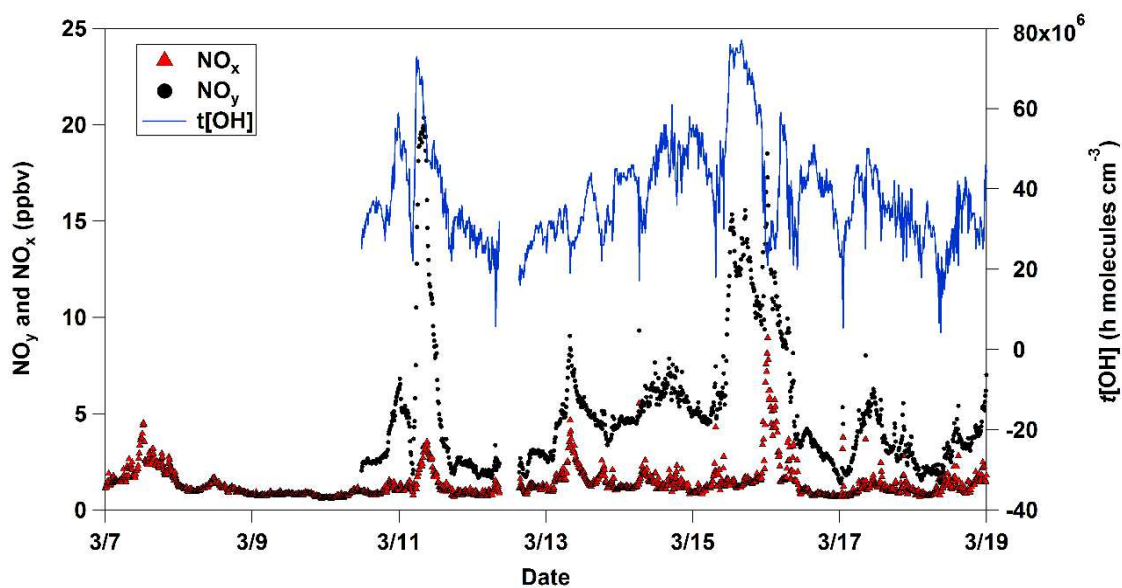

**Figure 6. Time series plot of observed $NO_x$ and $NO_y$ mixing ratios and estimated $t$[OH] at Fukue station.**

### Oxidation indicators: $\delta^{13}C$, $f_{44}$, and $t$[OH]

A plot of the daily averaged $\delta^{13}C$ of LV-WSOC versus the daily averaged $f_{44}$ of OA displayed a random variation for Fukue (Figure 7), which differed from the observations in 2010, implying an influence of the burning of $C_4$ plants.[3] A plot of $f_{44}$ versus $t$[OH] provides details of this difference (Figure 8): in this study $f_{44}$ ranged from 0.17 to 0.21 but did not increase proportionally with $t$[OH], while the $f_{44}$ ranged from 0.19 to 0.28 at Hedo, and increased proportionally with $t$[OH] ($r^2 = 0.81$) in the previous study.[3] The observations here can be explained by the saturation of $f_{44}$, indicating the contribution of a single component to the OA, such as HULIS. The saturation in turn indicated that the $f_{44}$ does not work as an oxidation indicator in this study. Although this explanation is consistent with the results from the PMF analysis discussed earlier (i.e., a single composition of LV-OOA), the case dependence of $f_{44}$ functionality is controversial and needs to be explored in more detail in future studies.

A plot of $\delta^{13}C$ versus $t$[OH] showed a more consistent trend with our understanding of SOA: the $\delta^{13}C$ value of SOA systematically increased as the oxidation of its precursor proceeds (Figure 9). Overall the plots seem to be consistent with this; however, it cannot be concluded that $\delta^{13}C$ can be used as an indicator for SOA.

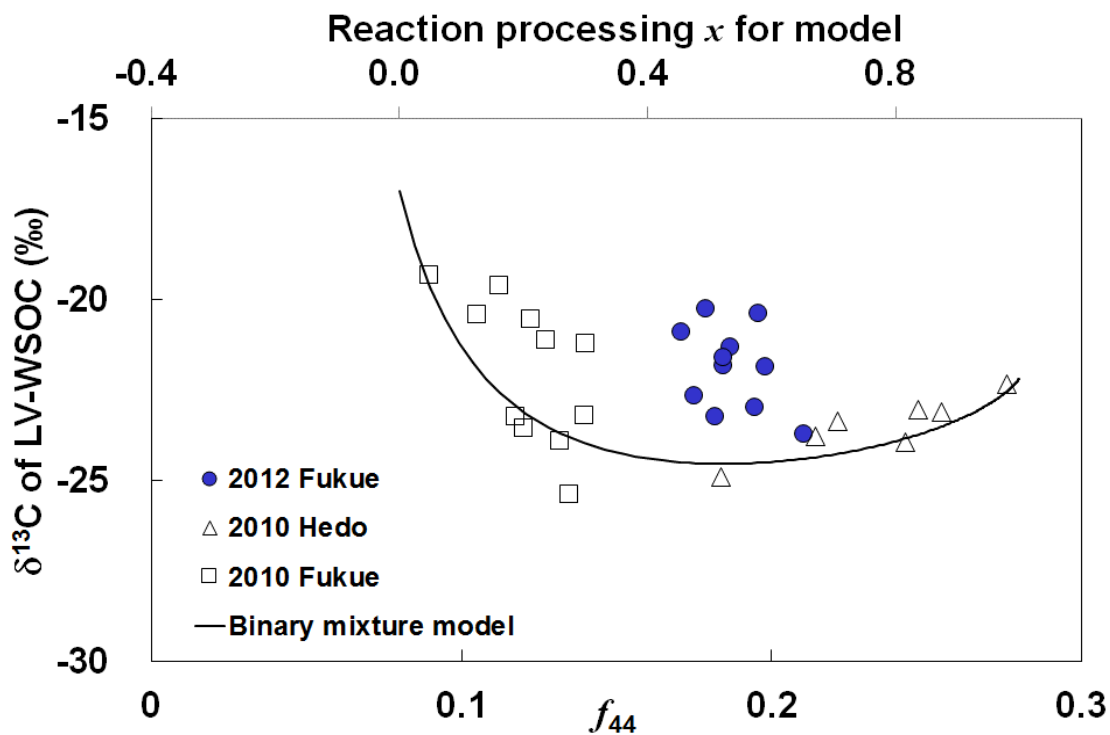

Figure 7. Plot for $\delta^{13}C$ of LV-WSOC versus $f_{44}$.

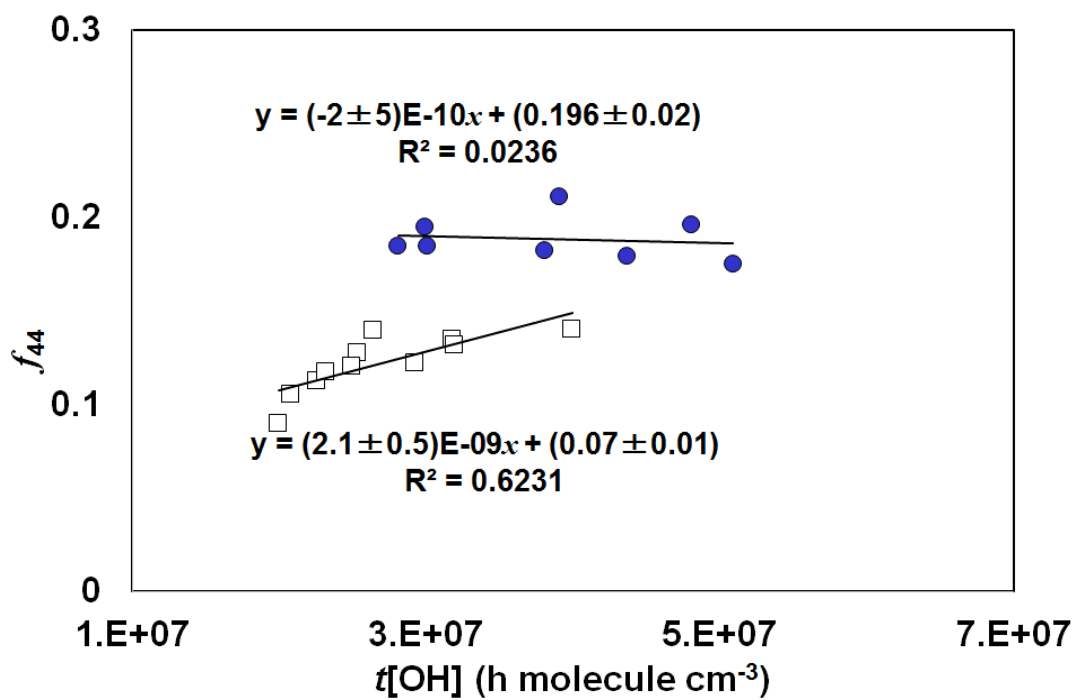

Figure 8. Plot of $f_{44}$ versus $t[OH]$ (24 h average) observed at Fukue in this study (blue circle) and in the December 2010 study (open square).

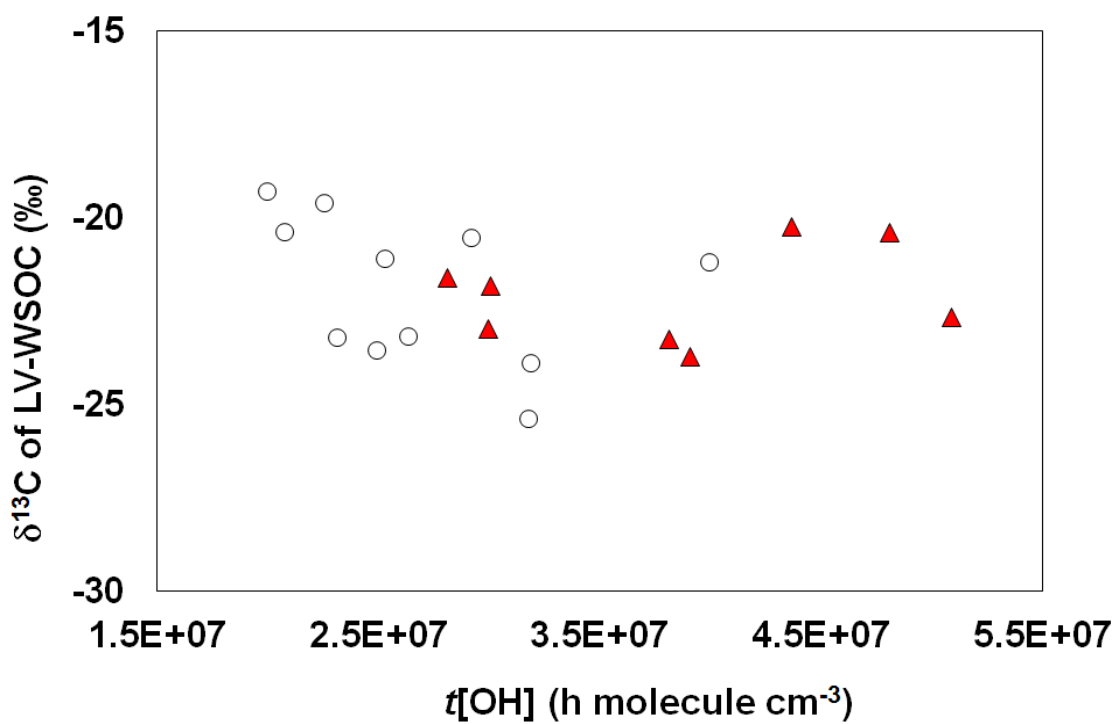

**Figure 9.** Plot for $\delta^{13}C$ versus photochemical age ($t$[OH]) estimated by $NO_x/NO_y$ ratio.

**Conclusion**

The observations in this field study showed a saturation of $f_{44}$ as measured by the ACSM, indicating a predominant composition of single compounds, such as HULIS. This composition may have produced a constant $f_{44}$; thus, it imposes a limitation on the use of $f_{44}$ as an oxidation indicator. The use of $\delta^{13}C$ as an oxidation indicator was expected to work under such circumstances; however, due to the large degree of scatter the association of $\delta^{13}C$ with $t$[OH] remains uncertain. Further studies of the SOA formed in the transboundary air during its transport over the East China Sea are required.

**Acknowledgements**
We thank Akio Togashi from the National Institute for Environmental Studies, and Ayako Yoshino and the students from Tokyo University of Agriculture and Technology for their help in the collection of filter samples, and for providing the sulfate concentration data. We also acknowledge the NOAA Air Resources Laboratory (ARL) for the provision of the HYSPLIT transport and dispersion model and/or READY website (http://www.ready.noaa.gov). This project was financially supported by the Internal Encouraging Research Fund of the Center for Regional Environmental Research at the National Institute for Environmental Studies, a Grant-in-Aid for Scientific Research on Innovative Areas (No. 4003) from the Ministry of Education, Culture, Sports, Science and Technology, Japan, the International Research Hub Project for Climate Change and Coral Reef/Island Dynamics of University of the Ryukyus, and the ESPEC Foundation for Global Environment Research and Technologies (Charitable Trust).